%% file: main.tex
\documentclass[11pt]{article}
\date{}

\usepackage{color,amsfonts}
\usepackage[urlcolor=blue,citecolor=black,linkcolor=black]{hyperref}
\usepackage{graphicx}
\usepackage{natbib}
\usepackage{amsmath}

\input{./Sections/macros}

\title{HypoSVI: Hypocenter inversion with Stein variational inference and Physics Informed Neural Networks}

\author{Jonathan D. Smith \footnote{Seismological Laboratory, California Institute of Technology, Pasadena, CA, USA}, Zachary E. Ross \footnotemark[1], Kamyar Azizzadenesheli \footnote{Lawson Computer Science Building, Purdue University, West Lafayette, IN, USA} , Jack B. Muir \footnotemark[1]}

\begin{document}

\maketitle

\begin{abstract}
\input{./Sections/0-Abstract.tex}
\end{abstract}

% --- Introduction ---
\input{./Sections/1-Introduction.tex}

\input{./Sections/2-Background.tex}

% --- Methods ---
\input{./Sections/3-Methods.tex}

\input{./Sections/4-Experiments.tex}

% --- Case Study ---
\input{./Sections/5-CaseStudy.tex}

% --- Discussion and Conclusions ---
\input{./Sections/6-DiscussionConclusions.tex}

\input{./Sections/acknowledgments.tex}

%====================================================
%====================================================
%====================================================
\bibliographystyle{agsm}
\bibliography{main}

\newpage
\input{./Sections/SupplementaryFigures.tex}

\end{document}

%% file: Sections/macros.tex
\newcommand{\rkhs}{\mathcal{H}}
\newcommand{\kernel}{\kappa}
\newcommand{\f}{\boldsymbol{f}}
\newcommand{\Sopt}{\mathcal{A}}
\newcommand{\Sdis}{\mathcal{D}}

%% file: Sections/0-Abstract.tex
We introduce a scheme for probabilistic hypocenter inversion with Stein variational inference. Our approach uses a differentiable forward model in the form of a physics informed neural network, which we train to solve the Eikonal equation. This allows for rapid approximation of the posterior by iteratively optimizing a collection of particles against a kernelized Stein discrepancy. We show that the method is well-equipped to handle highly multimodal posterior distributions, which are common in hypocentral inverse problems. A suite of experiments is performed to examine the influence of the various hyperparameters. Once trained, the method is valid for any seismic network geometry within the study area without the need to build travel time tables. We show that the computational demands scale efficiently with the number of differential times, making it ideal for large-N sensing technologies like Distributed Acoustic Sensing. The techniques outlined in this manuscript have considerable implications beyond just ray-tracing procedures, with the work flow applicable to other fields with computationally expensive inversion procedures such as full waveform inversion.

%% file: Sections/1-Introduction.tex
\section{Introduction}
Earthquake hypocenters represent the points in space and time at which earthquakes occur. They are a fundamental component of many downstream analyses in seismology, from seismic tomography to earthquake source properties. They are also used for real-time earthquake forecasting, such as during active sequences. Thus, the ability to reliably estimate hypocenters and characterize their uncertainty is of major importance in seismology.

Determining an earthquake hypocenter from travel time observations of seismic waves is a classic inverse problem in geophysics. The earliest methods performed linearized least-squares inversions, as described in \cite*{Geiger1910}. \cite*{Flinn1965} built on this work to include uncertainty into the theory through the adoption of confidence regions. Subsequent decades saw various adaptations of these methods to better account for uncertainty \citep{Buland1976,Bolt1970, Jordan1981,Uhrhammer1980,Jackson1979,Matsuura1984}. \cite*{Tarantola1982} proposed formulating the hypocenter inverse problem as one of Bayesian inference, allowing for complete probabilistic descriptions of the hypocentral parameters. The Bayesian treatment of this problem was expanded upon significantly \citep{Lomax2000,Lomax2005} as new techniques for statistical inference were developed, which allowed for more robust likelihood distributions to be used.

Bayesian inference  was pioneered by \cite{Jeffreys1935}, using Bayes rule to compute the posterior probability of whether the difference between two datasets is significant. Subsequent research in the geophysical community saw a wealth of understanding with an application of these methods to a geophysical setting \cite[e.g. ][]{Press1968}. In recent years, Bayesian inference has seen increasing adoption by the geophysics community for solving the numerous inverse problems that exist \citep{Hightower2020,Duputel2015,Gama2021}. One of the most popular classes of techniques is Markov Chain Monte Carlo (MCMC) sampling \cite[e.g. ][]{Gelman2013}, in which a target distribution is approximated by a series of samples drawn from it. Standard MCMC works well for low dimensional models with simple distributions, and has recently seen substantial improvement on larger dimensional ($\sim 1000$ parameters) models and more complex distributions \citep{Hoffman2011,Betancourt2017}. This has allowed for more widespread usage of MCMC (and thus Bayesian inference) in geophysics \citep{Fichtner2018,Fichtner2019}. One of the challenges with MCMC is that it often requires significant manual tuning of the sampling process to ensure convergence and mixing; another is that distributions with multimodal behavior can be difficult to sample from. An alternative class of techniques for Bayesian inference can be generally described as Variational Inference (VI) methods. With VI, the goal is to cast the inference problem as one of optimization, in which the target distribution is most commonly approximated by a parametric family of distributions. VI has seen some recent usage within geophysics research \citep{Nawaz2018,Nawaz2019,Zhang2020b}. One notable advancement in VI is an alternative called Stein variational inference \cite[SVI,][]{liu2016}, in which a collection of particles is iteratively optimized to approximate a target distribution. It is better suited than standard VI techniques at handling multimodal distributions, as the number of modes does not need to be known a priori, and does not require a parametric family of distributions to be assumed. The recent work of \cite*{Zhang2020b} outlined a method using SVI for full-waveform inversion, demonstrating that the SVI approach compared favorably with the results from Hamiltonian Monte Carlo \cite[HMC,][]{Hoffman2011}. In addition, \cite*{Zhang2020b} demonstrated that the SVI is highly parallelisable and could be more efficient than HMC for larger problems; however they found that the main bottleneck was the computational cost of the finite difference forward model simulations.

Recent advances in deep learning have led to the development of physics informed neural networks \cite[PINNs,][]{Rassi2019}, which are designed to learn solutions to partial differential equations (PDEs). Such approaches have a number of appealing properties that are not present with conventional approaches like finite difference methods \citep{Eaton1993}; for example the solutions can be made differentiable, are often mesh-free, and can be rapidly calculated upon demand. These properties make PINNs well-suited to be used as a forward model for solving inverse problems since it often is desirable to take gradients of an objective function, which is indeed the case for SVI.

Our contributions, as described in this paper, are as follows: (1) we develop a framework for earthquake hypocenter inversion using Stein variational inference; (2) we incorporate a PINN trained to solve the Eikonal equation as a forward model; (3) we perform experiments on the hyperparameters of the inverse problem to characterize their effect on the solution; and (4) we benchmark the method against a catalog of earthquakes from Southern California.

%% file: Sections/2-Background.tex
\section{Background}

\subsection{Stein Variational Inference}\label{Sec:SVI}

Variational inference (VI) is a class of approximate Bayesian inference techniques where, rather than sampling from a target distribution, the problem is cast as one of optimization. For two random variables $x$ and $y$, let $p(x)$ denote the prior on $x$, $p(y|x)$ the likelihood function, and $p(x|y)$ the posterior over $x$ after observing (conditioning on) $y$. Using Bayes rule, these quantities are related as, $p(x|y) \propto p(y|x)p(x)$. In standard VI, a flexible family of parametric distributions dependent on the parameters $\boldsymbol{\lambda}$, $q(x, \boldsymbol{\lambda})$, is used to approximate $p(x|y)$. Let the Kullback-Leibler divergence between $p$ and $q$ be defined as,
\begin{align}\label{eq:KLD}
    \Sdis_{KL}(q(x, \boldsymbol{\lambda}),p(x|y)) := \int_{-\infty}^{\infty} q(x, \boldsymbol{\lambda}) \log \left(\frac{q(x, \boldsymbol{\lambda})}{p(x|y)}\right) dx.
\end{align}
In essence, $\Sdis_{KL}$ measures how different two distributions are, with a value of 0 indicating $p$ and $q$ are identical. In typical VI problems, $\Sdis_{KL}$ is minimized with respect to the parameters of $q(x, \boldsymbol{\lambda})$. In the simplest scenarios, $q(x, \boldsymbol{\lambda})$ might take the form of a multivariate Gaussian distribution, and thus the goal is to learn the mean vector and covariance matrix. This all assumes that the family of distributions $q(x, \boldsymbol{\lambda})$ is even able to reasonably approximate $p(x|y)$.

An alternative way to perform VI is by using non-parametric estimates of $p(x|y)$. One such approach is termed Stein Variational Inference (SVI), in which $q(x, \boldsymbol{\lambda})$ is taken to be a collection of Dirac delta functions.  A single delta function minimizing $\Sdis_{KL}$ would coincide with the maximum a posteriori (MAP) point of $p(x|y)$. Thus, with SVI, the problem statement then is to arrange the set of delta functions in proportion to $p(x|y)$, while still satisfying eq. \ref{eq:KLD}. The main challenge is to prevent the delta functions from collapsing to the same MAP point; this is solved in SVI by introducing a mechanism for repulsion.

We now provide a rigorous mathematical treatment of SVI. Let $\rkhs$ denote a reproducing kernel Hilbert space on the domain $x$, with a positive definite reproducing kernel $\kernel$, endowed with the inner product $\langle\cdot,\cdot\rangle$ and the norm $\|\cdot\|_\rkhs$. We further define $\rkhs^d$, as a set of multivalued functions, with $d$ values, with the corresponding norm $\|\cdot\|_{\rkhs^d}$, where for any $\f=[f_1,f_2,\ldots,f_d]\in\rkhs^d$ we have $f_i\in\rkhs~\forall i\in[1,2,\ldots,d]$.

For a function $\f\in\rkhs^d$, we define Stein's operator ,$\Sopt$, endowed with $\rkhs^d$ and $p$ as,
\begin{equation}
(\Sopt \f)(x)=\f(x)\nabla_x\log p(x)^\top+\nabla_x\f(x).
\end{equation}
In this equation, the first term on the right-hand side is an outer product, while the second term on the right-hand side is the gradient of a vector-valued function.

We further define a kernelized Stein's discrepancy between two distributions $p$ and $q$ using $\rkhs^d$ is as follows,
\begin{align}\label{eq:Sdis}
    \Sdis(q,p) := \max_{\f\in\rkhs^d~s.t.,~\|\f\|_{\rkhs^d}\leq 1} E_{x\sim q}\left[trace\left(\Sopt\f\left(x\right)\right)\right]^2.
\end{align}
This discrepancy equals zero when $p=q$. Fortunately, the maximization in Eq.~\ref{eq:Sdis} has a closed-form solution $\Sdis(q,p)=\|\f^*_q\|_{\rkhs^d}$ where $\f^*_q:=E_{x\sim q}[\Sopt \kernel(x,\cdot)]$ is the maximizer.

Now consider the Kullback–Leibler divergence between $q$ and $p$, i.e., $\Sdis_{KL}$. We aim to find a gradient direction $g$ (a function), such that $g$ maximally reduces $\Sdis_{KL}$. Using $g$, we can use gradient descent with learning rate $\alpha$ and update  $q\leftarrow q-\alpha g$ to reduce the $\Sdis_{KL}$, and make the $q$ closer to $p$. It is known that for the kernelized Stein's discrepancy, the direction $g\in\rkhs$ that provides the direction of maximal change is $g:=\f^*_q$~ \citep{liu2016}. In the following, we provide an update rule to update $q$ and approximate the posterior $p$ given observed data, moving in the negative direction of maximum change.

We represent $q$ with a set of particles, i.e., a collection of many delta Dirac measures. $\{x_i\}_{i=1}^n$ where $q$ approximates $p(x|y)$. In the following, we update $q$, and make it closer to $p$ by moving the particles. Therefore, for the update direction $\f^*_q=E_{x\sim q}[\Sopt \kernel(x,\cdot)]$, at each point $x$, we have,
\begin{equation}\label{eq:discrepancy}
\begin{aligned}
\f^*_q(x, \boldsymbol{\lambda})&=\sum_{i=1}^n\Sopt \kernel(x_i,x)\\
&=\sum_{i=1}^n[\kernel(x_i,x)\nabla_{x'}\log p(x')|_{x'=x_i}+\nabla_{x'}\kernel(x',x)|_{x'=x_i},
\end{aligned}
\end{equation}

with the updating rule given by, 
\begin{align}\label{eq:update}
    x_{i}^{l+1}\leftarrow x_i^{l} - \alpha_{l} \f^*_q(x_{i}^{l}).
\end{align}
Here $\alpha_{l}$ is the step size at the $l$th epoch. For the choice of kernel, we use the Radial Basis Function (RBF), $\kernel(x',x)= \exp(-\frac{1}{h}\|x-x'\|^2)$, with $h$ representing the width of kernel, for its empirical and universal approximation properties. As discussed above, the update in Eq.~\ref{eq:update}, updates $q$ (through updating the particles distribution) at each time step to make it closer to $p$ in the Stein's discrepancy sense. 

\begin{figure*}
    \centering
    \includegraphics[width=1.00\textwidth]{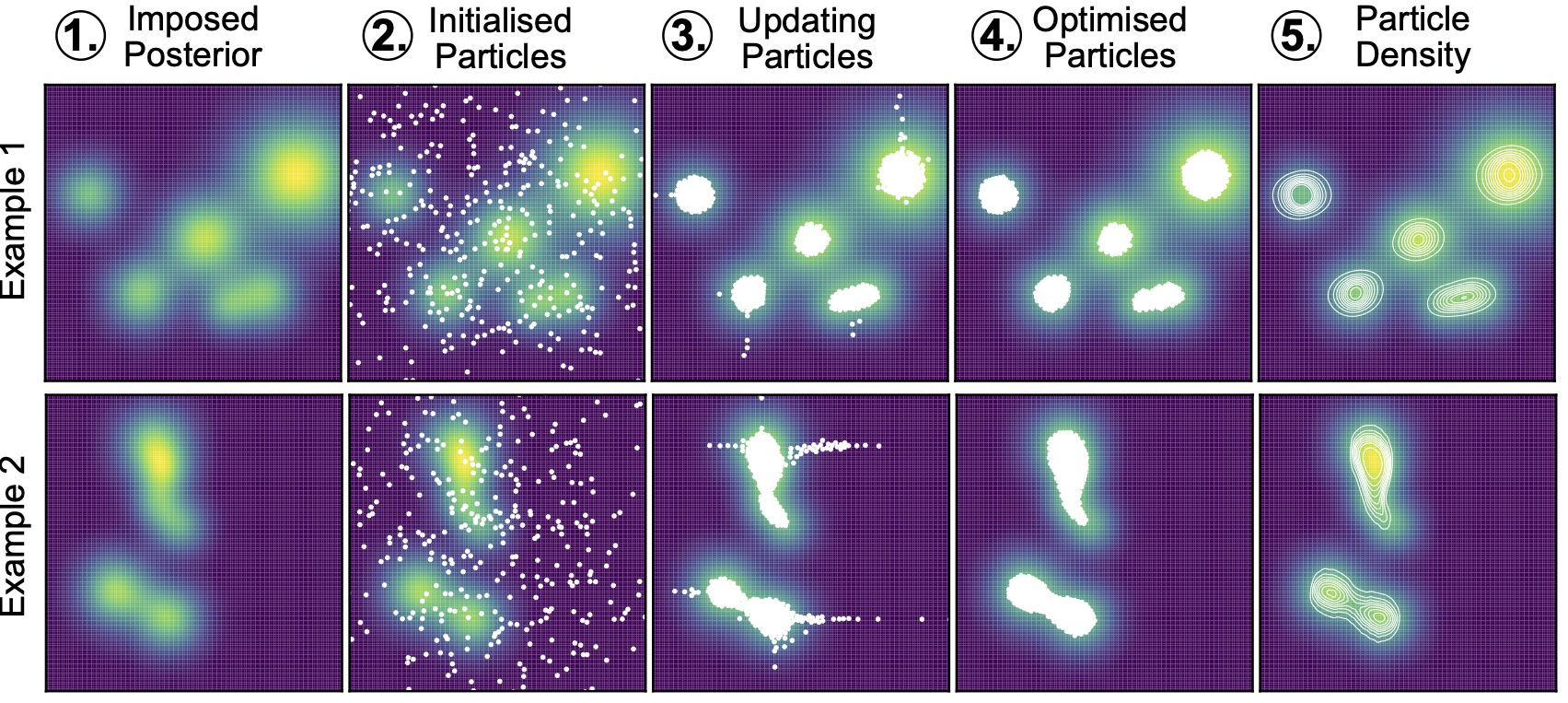}
    \caption{SVI applied to two examples distributions, each composed of a mixture of Gaussians. Stage 1 demonstrates the true posterior field. Stage 2 represents the initial randomised particle locations selected from a uniform prior. Stage 3 involves updating the particle locations to minimise the kernelized Stein discrepancy. By Stage 4, the optimised particle locations with locations have stabilized.}
    \label{fig:GaussianMix}
\end{figure*}

To give intuition for the SVI procedure, we applied it to two example distributions constructed from mixtures of Gaussians (Stage 1, Figure \ref{fig:GaussianMix}). Particle locations are first randomly selected within the domain using a uniform prior (Stage 2, Figure \ref{fig:GaussianMix}). The particle locations are updated by Equation \ref{eq:update} (Stage 3, Figure \ref{fig:GaussianMix}), with particles moving to minimise the kernelized Stein discrepancy. Once the particle locations remain static for multiple steps, the procedure terminates (Stage 4, Figure \ref{fig:GaussianMix}). The final density of the particles (shown by the contours in Stage 4 Figure \ref{fig:GaussianMix}) closely approximates the posterior distribution, demonstrating the validity of this method in approximating complex multimodal distributions.

\subsection{Physics informed Neural Networks for Ray Tracing}
In solving inverse problems for earthquake hypocenters, the most common approach is to use a ray theoretical forward model to calculate the expected travel times, $T$, for seismic waves propagating from a given source location to a receiver location. In heterogeneous 3D Earth models, the Eikonal equation is often solved to determine $T$ \citep[e.g.][]{rawlinson2005}, 
\begin{equation}
    \| {\nabla}_r{T_{s \rightarrow r}} \| ^2 = \frac{1}{V\left(\vec{x}_r\right)^2} = S\left(\vec{x}_r\right)^2 \label{eq:1}
\end{equation}
where $\|\cdot\|^2$ is the Euclidean norm, $T_{s \rightarrow r}$ is the travel time through the medium from a source location $s$ to a receiver  location $r$, $V_r$ is the velocity of the medium at the receiver location, $S_r$ is the slowness of the medium at the receiver location, and $\mathop{\nabla}_{r}$ the gradient at the receiver location.

The factored Eikonal formulation used throughout this methods mitigates the strong singularity effects at the source location by representing the travel time as a deviation from a homogeneous medium with  $V = 1$ \citep{treister2016}. The factored travel time form can be represented by:
\begin{equation}
    T_{s \rightarrow r} = T_0 \cdot \tau_{s \rightarrow r} \label{eq:2}
\end{equation}
where $T_0={\| \vec{x_r} - \vec{x_s} \|}$, representing the distance function from the source location, and $\tau$ the deviation of the travel time field from a model travel time with homogeneous unity velocity. Substituting the formulation of Eq. \ref{eq:2} into Eq. \ref{eq:1} and expanding using the chain rule, then the velocity can be represented by; 
\begin{equation}
\!V\!\!\left(\vec{x_r}\right)\! = \! \left[ T_0^2 \| \mathop{\nabla}_{r}\tau_{s\rightarrow r} \| ^2 \!+ \!2\tau_{s \rightarrow r}\left(\vec{x_r} - \vec{x_s}\right)\!\cdot\!\mathop{\nabla}_{r}\tau_{s \rightarrow r} + \tau_{s \rightarrow r}^2 \right]^{-\frac{1}{2}}\!\!\!\!\!\!. \label{eq:3}
\end{equation}
\cite*{Smith2020} developed a method to solve the factored Eikonal equation using physics informed neural networks (PINN), leveraging their inherent differentiability, without needing finite-difference solutions for training. Once fully trained, a network describing the travel time between any source-receiver pair can be represented by:
\begin{equation}
    T_{s \rightarrow r} = f_\theta\left(\vec{x_s},\vec{x_r}\right)
    \label{eq:fwdmodel}
\end{equation}
where $T_{s \rightarrow r}$ is the travel time between the source location $\vec{x_s}$ and receiver location $\vec{x_r}$, and $f$ is the neural network with weights and biases given by $\theta$. Gradients of the travel times are computed with automatic differentiation and used to determine the velocity at a set of receiver points. These "predicted" velocity values are compared with the user defined "true" velocity values at these same locations and used to define a misfit. The misfit is then minimized with respect to the parameters of the neural network. This results in a network that can rapidly determine the travel time between any two points within the user-defined 3D velocity volume. It should be emphasized that the neural network model is valid only for a single fixed velocity model; thus, changing the velocity model even slightly would require retraining of the neural network in some form. 

The PINN approach has several properties that are mathematically advantageous in solving inverse problems over conventional methods. First, the solutions to the Eikonal equation are mesh-independent, i.e. they are not discretised on a grid and can be evaluated at truly any point within the 3D medium. Second, the network is a forward model that is differentiable, allowing the user to rapidly determine the gradient of the travel time relative to the source and receiver locations. This also enables computing gradients of downstream objective functions, such as the recovered velocity of the network, seismic ray multipathing or hypocentral inversion. Third, by approximating the Eikonal equation with a deep neural network, the optimization part of the inverse problem is easily solved with graphical processing units (GPUs). This allows for the quick computation of travel times from the pre-trained model in addition to the higher order partial derivatives of the travel time relative to input terms; an imperative feature for the low computational cost of the inversion procedure in the later sections.

%% file: Sections/3-Methods.tex
% ==================================
\section{Methods}
% ---------------------------
\subsection{Overview}
\begin{figure*}
    \centering
    \includegraphics[width=1.0\textwidth]{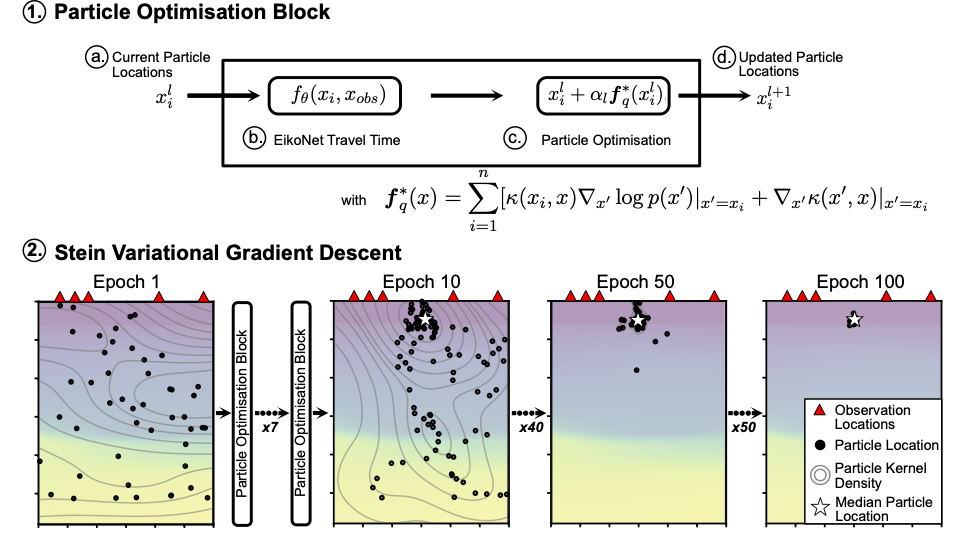}
    \caption{Overview of the inversion procedure. \textbf{Panel 1} the optimisation block that is applied to the particles to minimize the kernelized Stein discrepancy. (a) initial particle locations are supplied, (b) predicted travel times are determine from all particle locations to observation locations, (c) particle locations are then updated by a step in the direction that minimises the kernelized Stein discrepancy, (d) updated particle locations are returned. \textbf{Panel 2} demonstrates the optimisation scheme applying these optimisation blocks to update all the particle locations as a single batch between epochs. Red triangles represent observation locations. Black points represent particle locations. Contours represent the particle kernel density. White star represents the median location of the particles representing the optimal hypocentral location.}
    \label{fig:Overview}
\end{figure*}

We now present an approach for probabilistic hypocenter inversion that uses a PINN as a forward model and SVI to approximate the posterior distribution. The method consists of several primary steps:
    \begin{enumerate}
        \item An EikoNet model is trained for a given Earth velocity model to solve the Eikonal equation. This is performed for both P-waves and S-waves.
        \item A collection of particles is randomly initialized throughout the geographic study area. These represent preliminary hypocenter locations.
        \item Travel times are calculated with EikoNet from each particle to every receiver with an observation.
        \item The synthetic travel times are used together with the data to calculate a kernelized Stein discrepancy (loss function).
        \item The gradients of the loss are calculated with automatic differentiation and used to collectively update the particles' locations.
        \item Steps 3-5 are repeated until convergence. The final collection of particle positions will approximate the posterior distribution of the hypocenter.
        \item Uncertainty estimates are extracted from the particles by determining the percentile of the particle locations along each of the dimensions.
    \end{enumerate}
Next, we provide a detailed discussion of each stage of the procedure, with the outline of the inversion given in Figure \ref{fig:Overview}.

% ----------- Travel-Time formulation ----
\subsection{Constructing the forward model}\label{sec:EikoNetModels}
Throughout this study we train EikoNet travel time models using a set of constant training parameters and network architecture as described in Smith \textit{et al.} (2020) and supplied in Table \ref{Tab:EikoNetParams}. A model region is defined spanning our longitude, latitude, depth regional of interest, with xmin and xmax locations as [$117^o 30' W$, $32^o 30'N$, $-2km$] and [$115^o 30' W$, $34^o 30'N$, $50km$] respectively. The grid is projected to a UTM coordinate system, with random source-receiver locations selected within the UTM model space. These points represent the training locations, with different velocity models discussed below.

\begin{table}
\centering
\caption{EikoNet training paradigm used to learn velocity models}
\begin{tabular}{l|l}
\multicolumn{1}{l}{Parameter} & \multicolumn{1}{l}{Value}\\
\hline
Dataset Size & $1\times10^6$\\
Validation Fraction & $0.1$  \\
Batch Size & $752$\\
Optimizer & ADAM (+ scheduler)\\
Learning Rate & $1\times10^{-5}$ \\
Sampling Type & Weighted Random Distance\\
Sampling Type Bounds & $[0.1,0.9]$\\
Domain Normalization & Offset Min-Max Normalization \\
\hline
~~~Network Architecture & \begin{tabular}[c]{@{}l@{}} ~Dense $6\rightarrow32$ +
 Dense $32\rightarrow512$ \\
 \quad+ 
$10 \times$ Residual Blocks $512\rightarrow512$ \\
\quad+
 Dense $512\rightarrow32$ +
Dense $31\rightarrow1$ \\
~ELU Activation Function
\end{tabular}\\
\hline
\end{tabular}
\label{Tab:EikoNetParams}
\end{table}

\begin{figure}
    \centering
    \includegraphics[width=0.8\textwidth]{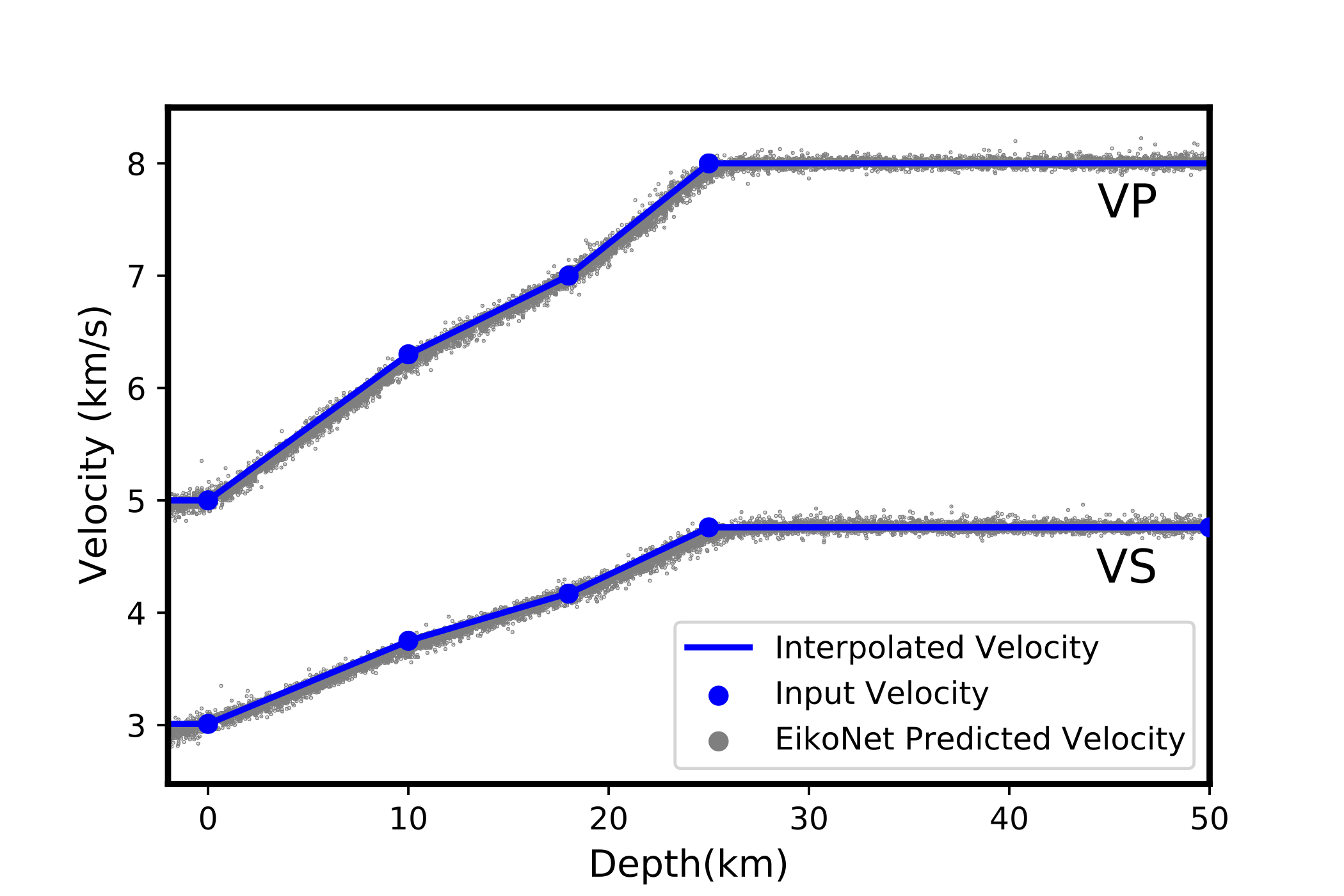}
    \caption{EikoNet trained travel time formulation for a one-dimensional velocity model only changing in depth (Z). Each curve represent a different model computed for both the P-wave velocity structure (VP) and S-wave velocity structure (VS). Blue points represent the user defined velocity values at depths, blue lines the linear interpolation of velocity between points. Gray points represent the predicted velocity from EikoNet for $1\times10^5$ randomly source-receiver pairs for each of the velocity models.}
    \label{fig:1DTravelTime}
\end{figure}

In many earthquake location procedures the complex geometry of the subsurface is poorly understood, with the assumption that lateral variations in velocity are negligible compared to velocity variations in depth. As such one-dimensional velocity structure describing how the velocity changes with depth are specified. These models typically have independent velocity structure defined for both the P-wave and S-wave arrivals, or a scaling relationship of Vp/Vs. It is important to understand how reliable these methods are for location procedures such as HypoSVI, as this would be a typical starting model for many use cases. In addition, understanding of the computational demand for training more simplistic travel time models, informs the feasibility of the method on typical computational systems. We investigate these problems for our region of interest by training EikoNet travel time models from the Vp and Vs velocity structure shown by the blue dots in Figure \ref{fig:1DTravelTime}a. We interpolate the velocity at the point locations as the linear interpolation of the observed velocity values. Two independent EikoNet neural networks are trained independently for the Vp and Vs velocity structure using the network parameters specified in Table \ref{Tab:EikoNetParams}. The training of each model took 10 epochs, with roughly a $10$ minutes training time on a Nvdia V100 GPU and $\sim 20$ minutes on a free Colab GPU (either a Nvidia K80,T4 or P100). Once trained the travel time models can be validated by comparing the imposed observed velocity to predicted velocity, determined as the analytical gradient of the travel time over the neural network, for a series of $1\times10^5$ source-receiver pairs within the three dimensional domain. Figure \ref{fig:1DTravelTime} outlines the comparison of the observed velocity structure and the predicted velocity, with the variance of the predicted velocity within $0.05km/s$ of the observed values. The consistent velocity structure and low computational overhead shows that this method is viable regardless of the available computational infrastructure.

\begin{figure*}
    \centering
    \includegraphics[width=1.0\textwidth]{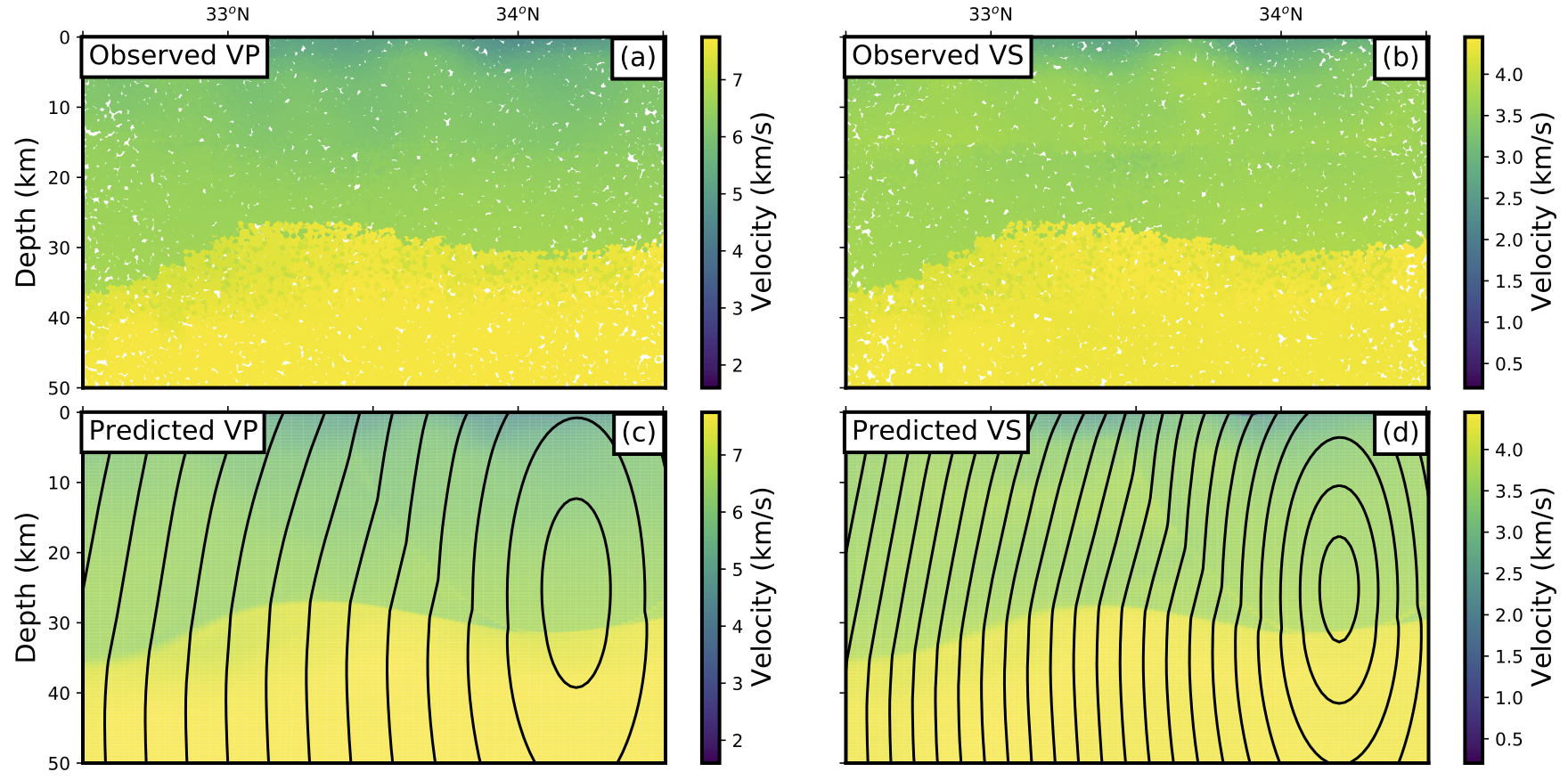}
    \caption{EikoNet trained travel time formulation for the complex three dimensional velocity of SCEC-CVM-H. Plots represent a slice in the three dimensional structure taken at longitude$=115^o30'W$. (a) and (b) represent the P-wave (VP) and S-wave (VS) velocity structures for the training points within $\pm 1.8'$ of the longitude slice and within the latitude and depth domain of the model space. (c) and (d) represent the predicted velocity structure colourmap and predicted travel time contours, at $2s$ intervals, for the P-wave and S-wave EikoNet models.}
    \label{fig:3DTravelTime}
\end{figure*}

For this study, our focus area is Southern California, and we use the SCEC-CVM-H velocity model \cite[version 15.1.0]{shaw2015}. We train EikoNet models to determine the travel time within the complex 3D velocity structures. The models are trained on $1 \times 10^6$ randomly selected source-receiver points within the domain, with example slice  at longitude$=115^o30'W \pm 1.8'$ given for the P-wave and S-wave velocity structure in Figure \ref{fig:3DTravelTime}a and \ref{fig:3DTravelTime}b respectively. The EikoNet models once trained represent the travel time and predicted velocity between any points, as such we show the recovered velocity model colourmap and travel time contours (at $2s$ spacing) for a earthquake source at $[115^o30'W,31^o12',25km]$ on a receiver grid as separation [latitude,depth] $= [0.05^o,0.5km]$ with longitude$=115^o30'W$. This example shows consistent agreement between the observed and predicted velocity models, able to reconcile the sharp velocity contrasts which create deflection in the travel time fields. This example demonstrates the viability of this method in complex 3D velocity structures.

% --------------------------
\subsection{Inverse problem formulation}\label{Sec:PosteriorDist}
An earthquake hypocenter, $m$, is composed of three spatial coordinates, $[x,y,z]$, and the origin time, $t_o$. Most commonly, the data used to locate earthquakes are measured times of seismic P- and S-wave arrivals ("phase picks") over a network seismic instruments. These phase picks define a set of absolute arrival time observations $d=T_{obs}$, where $d\in\mathbb{R}^N$. In a Bayesian framework \citep{Lomax2000}, inference on $m$ is performed by combining prior knowledge together with the observations, 

\begin{equation}
p(m|d) \propto p(d|m)p(m)
\end{equation}
where $p(m|d)$ is the posterior distribution, $p(m)$ is the prior distribution, and $p(d|m)$ is the likelihood. 

A simple example of $p(d|m)$ for hypocenter inversion is,
\begin{equation}
p(d|m) = \exp\left(-\frac{1}{2} \sum_{obs_i}\frac{\left[T_{obs}-T_{pred}\right]^2}{\sigma_i^2}\right)
\label{PDFGaussian}
\end{equation}
where $\sigma_i$ is an estimate of uncertainty and,
\begin{equation}
    T_{pred}=t_o+f_\theta\left(\vec{x_s},\vec{x_r}\right),
\end{equation}
is a nonlinear forward model, i.e. a solution to the Eikonal equation plus the origin time. Thus, the forward model in this problem is a PINN. Since $\vec{x_s}$ is included as an input to the neural network, this allows for downstream gradients to be taken with respect to it.

More recently, a likelihood function based on the Equal Differential Time method \cite[EDT]{Lomax2000} has seen increasing usage. The EDT likelihood builds differential times from all pairs of phases, and in the process, decouples origin time, $t_o$ from the spatial coordinates of the hypocenter. The formulation is given by;

\small
\begin{equation}
p(d|m) = \left[\sum_{a}\sum_{b}\frac{1}{\sqrt{\sigma_a^2+\sigma_b^2}}\exp\left(A\right)\right]^N,
\label{EQ:PDFEDT_1}
\end{equation}

\begin{equation}
    A = -\frac{\left[\left(T_{obs(a)}-T_{obs(b)}\right)-\left(T_{pred(a)}-T_{pred(b)}\right)\right]^2}{\sigma_a^2+\sigma_b^2},
    \label{EQ:PDFEDT_2}
\end{equation}
where $a$ and $b$ are different phase arrival time observations, $\sigma$ is a phase dependent estimate of uncertainty, and $N$ is the total number of differential times. In addition to reducing the number of free parameters by one (the removal of the origin time), this formulation acts to minimise the effects of outliers, which are particularly common with automated picking algorithms. This robustness results from the fact that in the EDT likelihood, the errors are combined in an additive manner. Each term in Eq. \ref{EQ:PDFEDT_1} produces a hyperbolic error surface that decays like a Gaussian in the direction normal to each point on the hyperbola. Thus, Eq. \ref{EQ:PDFEDT_1} can be viewed as producing a stack of hyperbolas with relatively limited intersection, which creates robustness in the presence of strong outliers.  However, the downside is that it results in posterior distributions that are highly multi-modal, making MCMC methods and standard variational inference schemes difficult to use for this problem \cite{Lomax2000}. To reduce the amplitude of the multimodal distribution of the posterior we can instead formulate the log-likelihood as that of a Laplacian differential time likelihood function. This represents the posterior space as a stacking of bands instead of hyperbolic surfaces. The equation then takes the form:
\begin{equation}
p(d|m) = \sum_{a}\sum_{b}\sqrt{2}\left|\frac{\left(T_{obs(a)}-T_{obs(b)}\right)-\left(T_{pred(a)}-T_{pred(b)}\right)}{\sqrt{\sigma_a^2+\sigma_b^2}}\right| + log\left( \frac{1}{\left(\sigma_a^2+\sigma_b^2 \right) \sqrt{2}}\right) 
\label{EQ:PDFEDT_1}
\end{equation}
Throughout this manuscript for all the testing we will use this Laplacian differential time likelihood function.

The origin time is reintroduced by using the optimised earthquake location to determine the predicted origin times to each of the observational locations, determining the origin time as the median of the predicted origin times. The uncertainty is then defined by the median absolute deviation (MAD) from the predicted origin time. We use a uniform prior, $p(m)$, with samples selected within the model domain specified in the Eikonal physics informed neural network.

The uncertainty in the posterior distribution is assigned as a combination of the observational, $\sigma_{obs}$, and forward model uncertainty, $\sigma_{pred}$, given as 
\begin{equation}
    \sigma^2 = \sigma_{obs}^2 + \sigma_{pred}^2.
\end{equation}
The observational uncertainty represents uncertainty in each of the observational times, with an expected standard deviation for each observation time supplied by the user. This value is then converted to a variance to define  $\sigma_{obs}$ for each observation.
The forward model uncertainty is constructed as a function of the predicted travel time for each of the observational locations \citep[similar to that given in ][for LOCGAU2]{Lomax2000}, given by
\begin{equation}
\sigma_{pred}=\begin{cases}
			\sigma_{min}, & \text{for $\sigma_{f}T_{P} < \sigma_{min}$}\\
            \sigma_{frac}T_{pred}, & \text{for $\sigma_{min} \leq \sigma_{f}T_{P} \leq \sigma_{max}$}\\
            \sigma_{max}, & \text{for $\sigma_{f}T_{P} > \sigma_{max}$}\\
		 \end{cases}
\end{equation}
where $\sigma_{f}$ is the fraction of the travel time to use as uncertainty, bounded within the max and min uncertainties specified by $\sigma_{min}$ and $\sigma_{max}$ respectively. Throughout this work we use the $[\sigma_{f},\sigma_{min},\sigma_{max}] = [0.1,0.1s,2.0s]$, discussing the effects of these parameters on synthetic testing within Section \ref{Sec:ObsWeight}.

 A SVI procedure is used together with the Laplacian differential time likelihood function. We use a RBF kernel, for its practical and universal approximation properties. First, we initialize $N$ particles randomly using a uniform prior over the 3D study area. For each of these particle locations, we calculate corresponding travel times using EikoNet forward model, evaluating the posterior (to within the normalization constant), and determine the kernelized Stein discrepancy. Then, we calculate the gradients of this loss function particle-wise with respect to the hypocentral coordinates using automatic differentiation, which is possible due to the differentiability of the PINN. We use these gradients together with the ADAM optimizer \citep{kingma2014} to update the particle locations until convergence, where the optimal hypocentral location is consistent across multiple epochs. The optimisation is stopped using an early stopping criterion, where the particle locations are consistent for at least 5 epochs Supplementary Video 1 demonstrates the convergence for the example outlined in Figure Figure \ref{fig:Overview}.

The next step is to extract summary statistics from the posterior distribution. We determine the 95\% credibility interval of the particle locations in each of the Cartesian dimensions and take this as the uncertainty in the earthquake location. All particles particle locations are returned for each earthquake, for additional high level statistical analysis.

%% file: Sections/4-Experiments.tex
\section{Experiments}\label{sec:ParamOpt}
\subsection{Method validation}

\begin{figure*}
    \centering
    \includegraphics[width=1.0\textwidth]{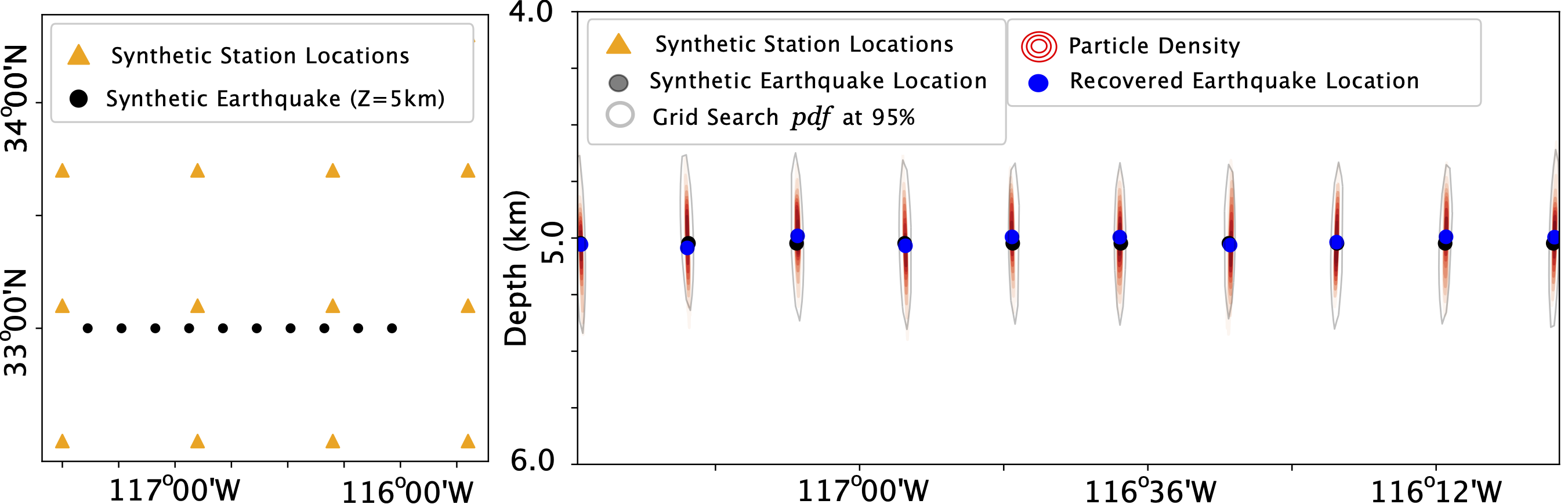}
    \caption{Synthetics earthquake location recovery for a synthetic seismic array. (a) represents a map view of the synthetic earthquake locations and synthetic stations locations. Black points represent the synthetic earthquake location latitude and longitudes, at a fixed depth of $5km$. Yellow triangles represent the synthetic station locations, at a fixed depth of $0km$. Black line represent a cross section at a fixed latitude, with the cross section given in (b). (b) represents the true earthquake locations, black points, recovered optimal location, blue dots, posterior determined by the particle density, red contours, and grid search derived posterior at 95\%, gray line.}
    \label{fig:OverviewSynthetics}
\end{figure*}

In this section we first demonstrate the earthquake inversion scheme on a series of synthetic tests.  We construct a catalogue of synthetic earthquake locations across the region, determining the travel time to a grid of observation points at fixed elevation of $0km$, before applying a $0.01s$ uncertainty in the synthetic phase arrival and inverting to determine the earthquake location and uncertainty. The earthquake locations are at a fixed latitude and depth of $33^o00'N$ and $5km$ respectively, with longitude varying from $117^o23'W$ to $116^o57'W$ at $9'$ separations. The recovered optimal hypocentre and location uncertainty are then compared with the true earthquake locations and an expected 95 percentile contour from a the solution of a grid-search inversion. We vary the possible user defined parameters with the optimised parameters given in Table \ref{Tab:HypoSVIParams} and earthquake locations in Figure \ref{fig:OverviewSynthetics}. However, we expect that these parameters will need to be varied somewhat depending on the exact application, for example if the error models or network geometry are changed significantly. As such we recommend that initial synthetic testing is undertaken before real data is inverted. Outlined below are discussions on how each hyperparameter affects the recovered locations for this study, with corresponding Supplementary Figures \ref{Supfig:Params_Particles}-\ref{Supfig:Params_VeloMod}.

\begin{table}
\caption{HypoSVI parameters used in earthquake location techniques}
\centering
\begin{tabular}{l|l}
\multicolumn{1}{l}{Parameter} & \multicolumn{1}{l}{Value} \\
\hline
Number of Particles & $150$\\
Observation Weights & $[0.1,0.1s,2.0s]$\\
Radial Basis Function & $15$\\
\hline
\end{tabular}
\label{Tab:HypoSVIParams}
\end{table}

\subsection{Number of particles}\label{Sec:NumParticles}
The number of particles used in SVI is of great importance for the resolution of the resolved earthquake location. If the number of particles is too small then the particles density is unable to adequately represent the posterior distribution. However, a large number of particles would have increasing computational demand on the inversion procedure and is intractable for large earthquake catalogues. We specify a optimal number of samples equal to $150$ and find that an increase in the number of particles does not provide additional information on the the earthquake location, but reducing the number of particles greatly (e.g. 10 particles) effects posterior. Additional plots for variations of number of particles, with remaining parameters set equal to Table \ref{Tab:HypoSVIParams}, is given in Supplementary Figure \ref{Supfig:Params_Particles}.

\subsection{Influence of the kernel}
The RBF kernel can be represented by $\kernel(x,x')= \exp(-\frac{1}{h}\|x - x'\|^2)$, where $h$ is the shape parameter and $x$ the pairwise particle difference. The second term in the kernelized Stein discrepancy (Eq. \ref{eq:discrepancy}) represents the gradient of the kernel, acting as a repulsive force to prevent all the particles from collapsing into local modes. This term reduces to $\sum_i \frac{2}{h}(x-x_i)k(x_i,x)$ that drives $x$ from the neighboring $x_i$ that have large $k(x_i,x)$. Understanding the trade off for the shape parameter is important as larger values could affect the recovered posterior. \cite{liu2016} defined a dynamic shape parameter with the value changing depending on $h=\text{med}^2/\log n$, where $\text{med}$ is the median distance between pairwise particles. The definition $\sum_j k(x_i,x_j) \approx n \exp(-\frac{1}{h} \text{med}^2) = 1$ demonstrates that for each $x_i$ the contribution from its own gradient and the influence from other points balances out. We investigate the variation of the RBF shape parameters on the recovered synthetic earthquake locations finding that increasing the parameter acts to drive particles further away, decreasing the particle density close to the optimal recovered hypocentral location (Supplementary Figure \ref{Supfig:Params_RBF}). We decided to use a static shape parameter of $15 km$, to mitigate any difference that could occur to the posterior from multiple runs of the same observations for a dynamic shape parameter. We attribute this parameter as a user defined variable that should be calibrated for the regional context of interest, expecting the optimal hypocentral location to not vary that much but the returned location uncertainty to increase with larger RBF values.

\subsection{Error models}\label{Sec:ObsWeight}
The total uncertainty assigned to the inverse problem is a combination of the picking uncertainty and the forward model uncertainty due to the velocity structure. As described previously, we follow \cite{Lomax2000} and characterize the uncertainty in the forward model as a fraction of the travel time. This is a reasonable choice as the uncertainty in the predicted travel times is expected grow in proportion to the travel time. In our hyperparameter investigation we found that a fraction of $0.1$ should be used, as lower values lead to significant mis-location of the recovered events (Supplementary Figure \ref{Supfig:Params_VeloMod}). The upper and lower bounds to the allowed error has less of an effect on our synthetic testing, which we attribute to the synthetic station locations being regularly spaced. For observational data that is clustered spatially the upper and lower bounds could be of great importance and should be investigated with synthetic examples for the specific network geometry.

\subsection{Computational demands}\label{sec:ComputDemand}
The number of observations going into a inversion affects the compute time, as each observation requires predicted travel time formulations from EikoNet and gradients to be computed . Here, we investigate the computational cost of the inversion procedure while increasing the number of observations. We replicate an increasing number of observations by copying the synthetic station deployment locations multiple times, labelling them as different station names but comprising the same arrival times. This synthetic testing was chosen to minimising the changing effect on the location estimate, which would occur if additional synthetic station locations are provided. All other location hyperparameters are fixed at values given in Table \ref{Tab:HypoSVIParams}. The earthquake locations are then determined for the varying number of observations and the total number of pairwise differential times, with the average computational time for a Nvidia V100 shown in Table \ref{Tab:HypoSVICompCost}. The computational time even for the $2048$ observations, $2055378$ differential times, only takes $439s$ per event. These synthetic tests demonstrate that this approach is computationally scaleable with computational time increasing as a linearly in a log-log space of computational time vs number of observational differential times.

\begin{table}
\caption{HypoSVI computational cost on a Nvidia V100 GPU with different number of observations and corresponding differential time pairs. The remaining parameters used in this synthetic test are given Table \ref{Tab:HypoSVIParams}}
\centering
\begin{tabular}{l|l|l}
\multicolumn{1}{l}{\# of Observations} &
\multicolumn{1}{l}{\# of Differential Times} &
\multicolumn{1}{l}{Time per Event(s)}\\ \hline
$32$  &$496$    & $6$\\
$128$ &$8128$   &$17$\\
$512$ &$130816$ &$64$\\
$1024$&$523776$ &$155$\\
$1408$&$990528$ &$247$\\
$1728$&$1492128$&$336$\\
$2028$&$2055378$&$439$\\
\hline
\end{tabular}
\label{Tab:HypoSVICompCost}
\end{table}

%% file: Sections/5-CaseStudy.tex
\section{Case Study: Application to earthquake swarms in Southern California}
\subsection{Background}
To further validate the developed method, we apply it to real earthquakes occurring within the Southern California region, with region defined in Section \ref{sec:EikoNetModels}. This study area was chosen as it encompasses a large seismic network and complex 3D regional velocity structures \citep{allam2012}. We used the detections and phase picks from the open source Southern California Earthquake Data Centre (SCEDC) phase arrival observational catalogue, for the fist $10k$ events starting 2019-01-01. The events and phase picks used have all been manually reviewed by analysts at the Southern California Seismic Network \citep{hutton2010}.

% --------------------------------------------------------
\subsection{Earthquake Location comparisons with NonLinLoc}
\begin{figure*}
    \centering
    \includegraphics[width=1.0\textwidth]{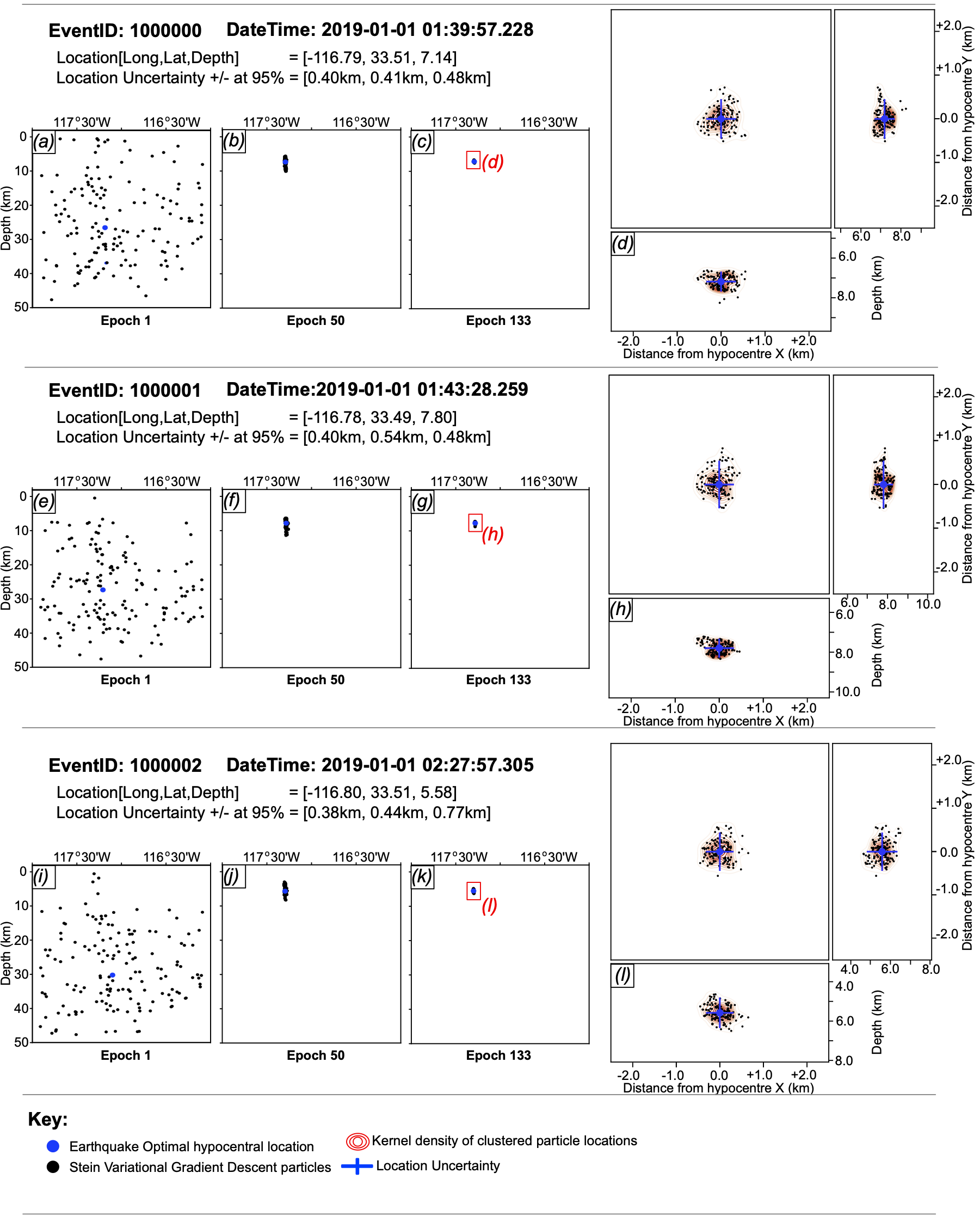}
    \caption{Example earthquake locations for three earthquakes in the Catalogues using travel-times derived from the three-dimensional regional velocity model. Left panels represent the particle locations changing at different epochs in SVI. Right panels represent a zoom in of the final event locations, with the particle locations shown relative the recovered optimal hypocentral location. Kernel density contours are shown in red for the clustered particles.}
    \label{fig:ExampleEarthquakes}
\end{figure*}

We infer hypocenters for the $10k$ earthquakes using two different velocity models (1D and 3D cases, described in Section \ref{sec:EikoNetModels}). The hyperparameters used for the inversions are outlined in Table \ref{Tab:HypoSVIParams} with detailed explanation of the reasoning behind the parameter definition outlined in Section \ref{sec:ParamOpt}. The catalogues are generated on a Nvidia V100 GPU with an average of $5s$ per event, varying depending on the number of observations in the inversion procedure, with on average $\sim 30$ observations per event. Since the calculation of travel-times from EikoNet is independent on the complexity of the velocity model (once the network has been trained), the processing takes equal time for both the 1D and 3D trained models. Example inversions for three events are shown in Figure \ref{fig:ExampleEarthquakes}.\\

\begin{figure*}
    \centering
    \includegraphics[width=1.0\textwidth]{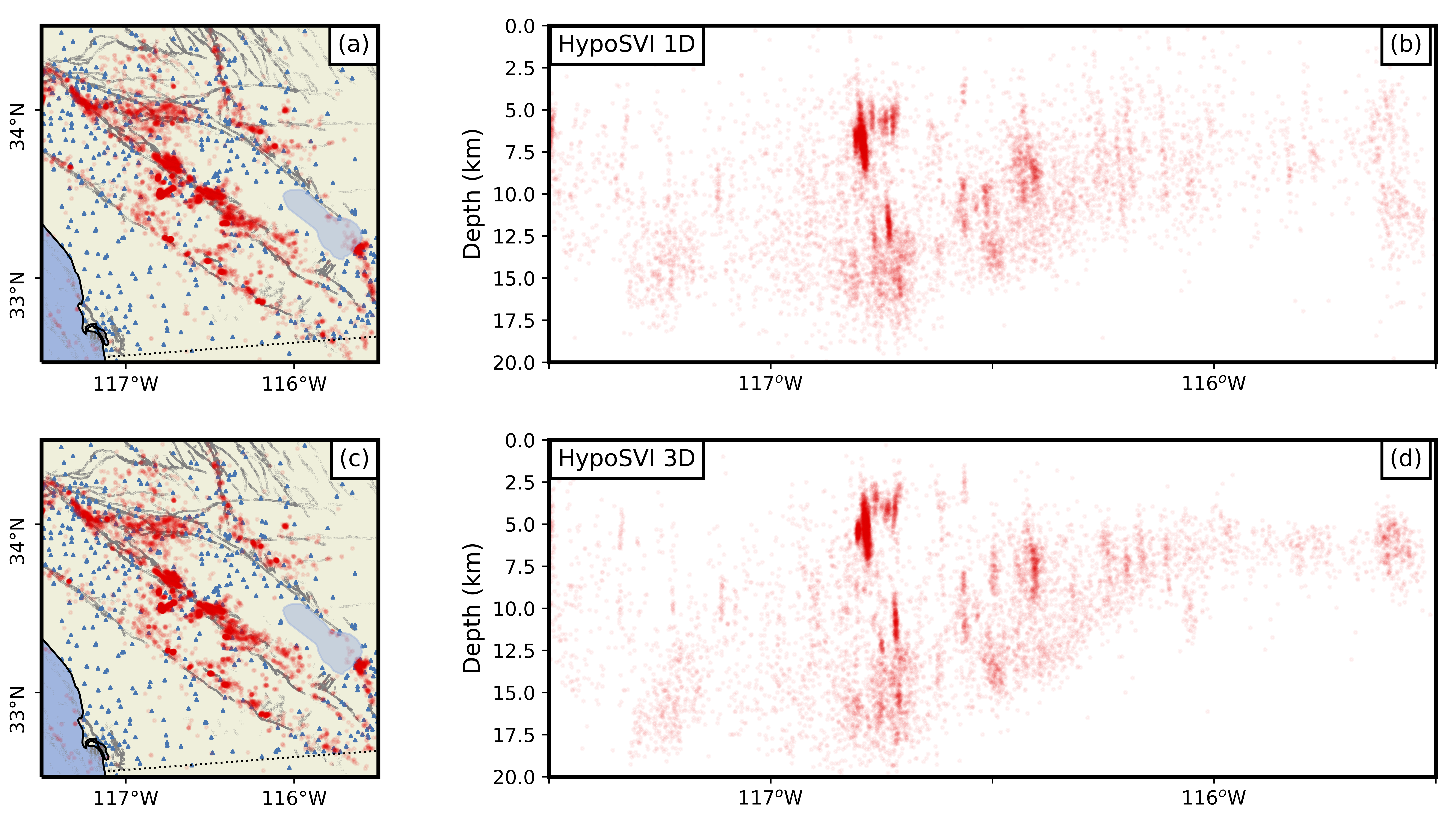}
    \caption{Comparison of earthquake locations between the HypoSVI locations using a regional 1D velocity and SCEC-CVM-H 3D velocity structure. Left column represents the latitude/longitude map of the detected earthquakes given by red dots, observational station locations given by blue triangles and mapped faults by gray lines. Right column represents a longitude vs depth cross-sections of earthquakes. (a) and (b) are the locations determined from HypoSVI with a EikoNet model trained on a regional 1D velocity. (c) and (d) are the locations determined from HypoSVI with a EikoNet model trained on a regional SCEC-CVM-H 3D velocity structure.}
    \label{fig:EarthquakeLocs}
\end{figure*}

To understand the validity of our location technique we compare our earthquake catalogue, with a catalogue determined using the conventional earthquake location software, NonLinLoc. NonLinLoc is a non linear earthquake technique leveraging finite-difference travel-time solutions; Gaussian or equal-differential likelihood functions; and, likelihood estimations schemes using oct-tree, grid-search or Markov Chain Monte Carlo (MCMC). Travel-times are computed by solving the eikonal using a finite-difference approach  outlined in \cite{podvin1991}. For a 1D velocity structure, only varying in depth, the package computes the travel-times as an radial 2D finite-difference travel-time model that depends on the radial distance from the observation point and the depth, saving these as independent travel-time look-up tables. In contrast, for complex three-dimensional velocity structures the travel-times are computed for a user defined gridded series of receiver locations, with each observation saved as a separate travel-time  look-up table. Since the storage and computational requirements for a NonLinLoc using the complex 3D velocity for a very high resolution location grid, this method was intractable as it would return large gridding artifacts to the recovered earthquake locations and predicted location uncertainty, which are not directly comparable to the non-gridded solutions of the HypoSVI. Instead we compare the HypoSVI and NonLinLoc locations using the one-dimensional velocity structure, with the NonLinLoc travel-time and initial location grids resolved to $1$km  and $2$km receptively. The location is determined using a Equal-Differential Travel-Time (EDT) likelihood function and octree sampling technique. The location uncertainty of the recovered NonLinLoc catalogue is determined as the standard error in X,Y,Z to 2-std using the diagonal of the covariance matrix. The remaining NonLinLoc user parameters are given in the full control file in the Supplementary Material. The HypoSVI earthquake catalogues for the 1D and 3D velocity structures are given in Figure \ref{fig:EarthquakeLocs}a-b and \ref{fig:EarthquakeLocs}c-d respectively.\\

% -----------------

\begin{figure*}
    \centering
    \includegraphics[width=1.0\textwidth]{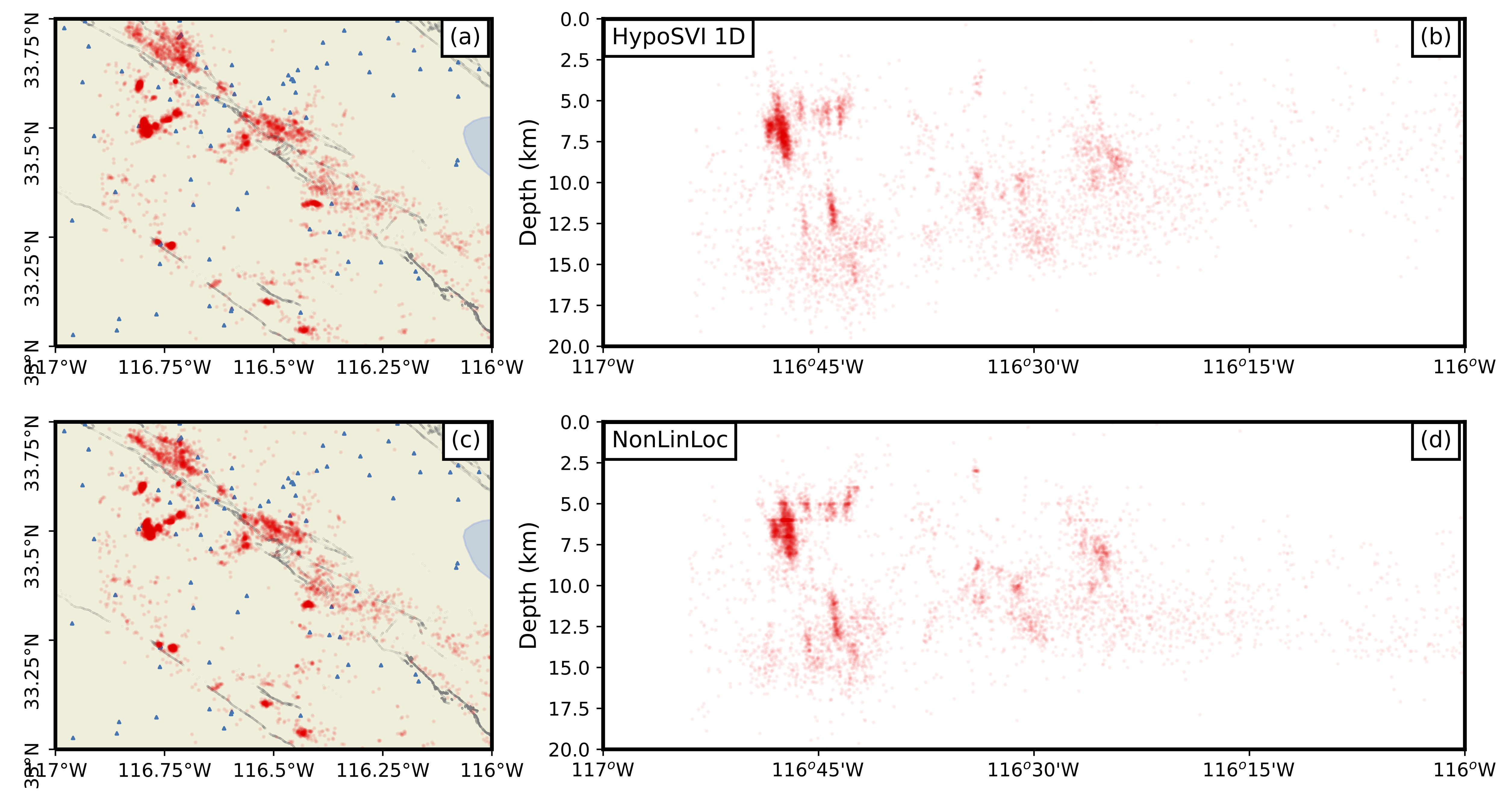}
    \caption{Zoom in earthquake location comparison for the region for subregion of [$117^oW$,$33^oN$] to [$116^oW$,$33^o45'N$]. (a)-(b) are the locations determined from HypoSVI with a EikoNet model trained on a regional 1D velocity. (c)-(d) are the locations determined from the NonLinLoc inversion procedure.}
    \label{fig:Map_NLLocSVI}
\end{figure*}

\begin{figure*}
    \centering
    \includegraphics[width=1.0\textwidth]{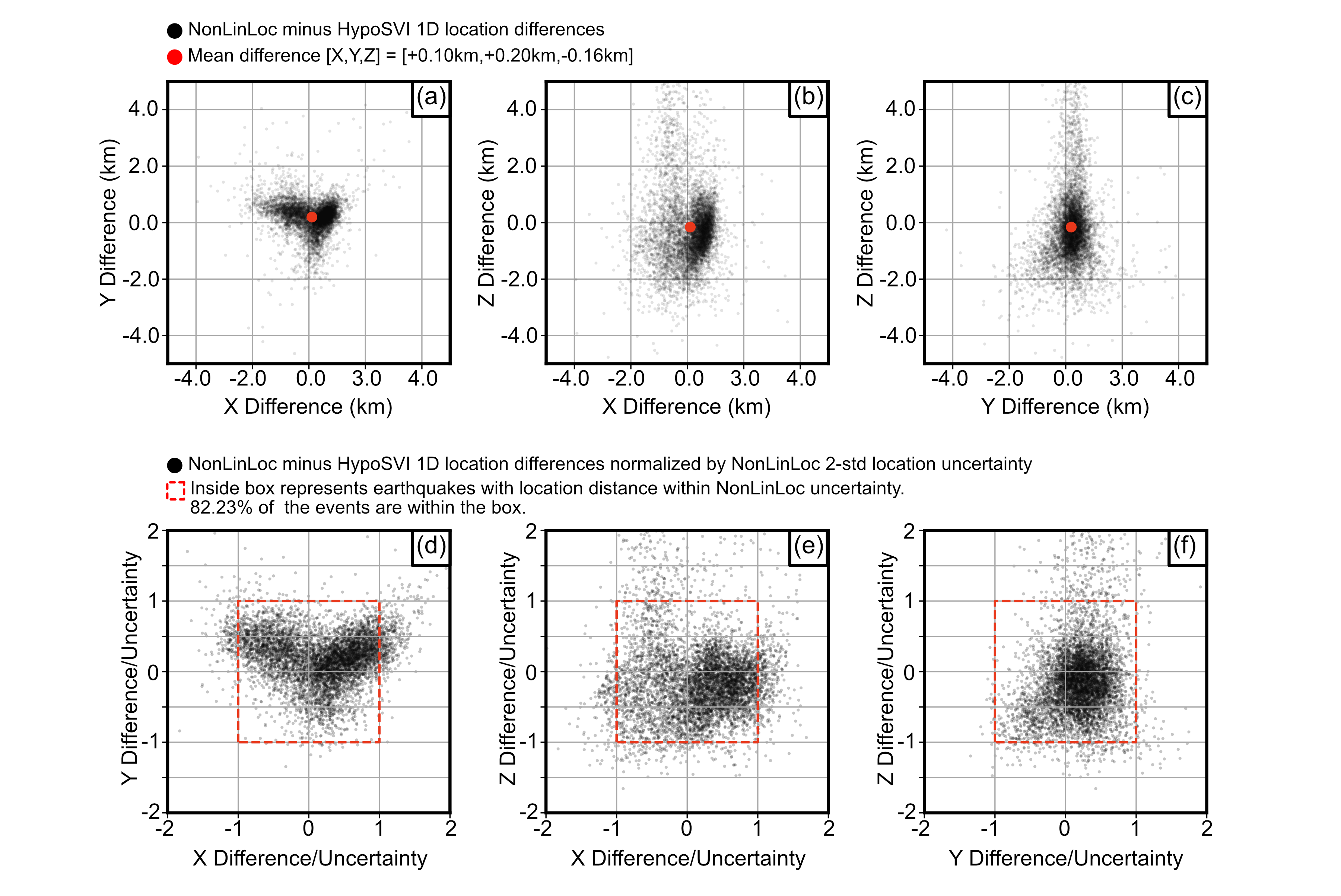}
    \caption{Earthquake distance comparison for the NonLinLoc and HypoSVI 1D catalogue for the region [$117^oW$,$33^oN$] to [$116^oW$,$33^o45'N$], projected to the local $X$,$Y$,$Z$ UTM coordinate system. (a)-(c) black dots represent the relative distance between catalogue event locations in X,Y,Z; with red dot representing the mean location. (d)-(f) black points relative distance between catalogue event locations normalized by the NonLinLoc 2-std location uncertainty. Red-dashed region represents the catalogue events with a relative distance less than the location uncertainty.}
    \label{fig:Sta_NLLocSVI}
\end{figure*}

For comparison we derive a NonLinLoc catalogue  for subregion of [$117^oW$,$33^oN$] to [$116^oW$,$33^o45'N$]. This region comprises a total of $6307$ events in the HypoSVI 1D catalogue (Figure \ref{fig:Map_NLLocSVI}a-b), with the NonLinLoc comprising $6383$ events (Figure \ref{fig:Map_NLLocSVI}c-d). Manual inspection showed that the events present in the NonLinLoc catalogue but not HypoSVI catalogue, are events that are locate external to the subregion in the HypoSVI catalogue but are projected to the edge of NonLonLoc search grid, having large location uncertainties. For the remaining events we determine the relative location differences between the two catalogues by projecting both catalogues to a local Universal Transverse Mercator (UTM) coordinate system and determining the distance between the events in $km$ in a local XYZ coordinates. The relative distance of the NonLinLoc locations minus the HypoSVI 1D locations are given in Figure \ref{fig:Sta_NLLocSVI}a-c. The relative locations demonstrate no consistent spatial bias, with the mean location difference given by $[X,Y,Z] = [+0.07\text{km},+0.19\text{km},-0.41\text{km}]$, as shown by the red dot in Figure \ref{fig:Sta_NLLocSVI}a-c. In addition, we normalise the location difference by the location uncertainty from the NonLinLoc catalogue. Figure \ref{fig:Sta_NLLocSVI}d-f gives the normalized location distances, with $83.29\%$ of the events having a relative distance less than that of the NonLinLoc location uncertainty, as shown by the points within the dashed box. Although there is similarity between the catalogues, some bias is observed in the NonLinLoc catalogue with many events displaying gridding artifacts attributed to a receiver grid required for the inversion procedure. In addition, this subregion was selected over the global dataset as the NonLinLoc procedure was computationally intractable across the whole region.

%% file: Sections/6-DiscussionConclusions.tex
\section{Discussion}
The experiments of the previous section demonstrate that the methodology developed in this paper is able to reliably approximate the posterior distribution for earthquake hypocenters from travel time observations by tuning the parameters $\boldsymbol{\lambda}$ of a parametric distribution to approximate the posterior. We believe that the general setup of the methodology could extend to other geophysical inverse problems such as tomography, however these other scenarios would of course require a differentiable forward model to efficiently compute the necessary gradients. 

The non-gridded earthquake locations obtained with HypoSVI demonstrate improvement over those derived with gridded schemes. This results from the continuous nature of the SVI procedure and forward model EikoNet. In addition, the HypoSVI results show similar earthquake locations for the regional 1D and SCEC-CVM-H 3D velocity model, but with the computational time independent of the velocity model provided. This has considerable advantages over methods that require a high resolution 3D travel-time model for each station, making the computational cost of the earthquake location inversion procedure intractable for the specific examples in this study.

Another advantage of our approach is that it is computationally efficient and can make use of state of the art GPU hardware and modern deep learning APIs like PyTorch. This allows for rapid calculation of the gradients with automatic differentiation. As GPU hardware improves, such as through increased memory, these performance gains will be passed on to the algorithm which will allow for even larger datasets to be worked with than currently possible. By combining SVI with EikoNet, we are able to evaluate observations at any point within the 3D volume without retraining, i.e. the forward model is valid for any array geometry. Due to the highly-parallelized nature of calculations with neural networks, our method scales well to very large networks, which may be important for emerging technologies like Distributed Acoustic Sensing (DAS). This was demonstrated herein by the ability to locate an earthquake with 2048 phase picks in 439 seconds. Thus, our HypoSVI approach is ideal for handling the enormous data volumes that are starting to emerge in seismology. 

The procedure outlined in this manuscript requires a pre-trained EikoNet travel time model for a user supplied subsurface velocity structure. This inherently assumes some prior knowledge of the sub-surface velocity structure, which could inherently be wrong, giving false solutions. In future work the HypoSVI procedure could be updated not only to optimise for the hypocentral locations, but also leverage transfer learning techniques to update the EikoNet for an improved velocity model for the region. For this idea, the earthquake locations would be initially determined from the initial velocity model. The EikoNet parameters would then be updated to minimise the misfit between the observed and predicted arrival times, by updating the velocity model using the adjoint-state method \cite{Sei1994}, while still satisfying the factored Eikonal equation. Finally the SVI particles from the initial earthquake locations would be updated using the current iteration of EikoNet. These procedures would be repeated several steps until misfit both for the travel time formulations and earthuqake location posteriors are minimised.

The current HypoSVI procedure is applied to individual events with no relative relocation between event pairs. The procedure could be expanded to a relative relocation scheme, using the relative travel time differences between events or even cross-correlation similarities between the events. This relative relocation approach could leverage the continuous travel time formulations from the EikoNet models.

\section{Conclusions}
In this paper, we developed a new approach to performing Bayesian inference on earthquake hypocenters that combines a differentiable forward model (physics-informed neural network) with SVI. Unlike with MCMC sampling methods, SVI approximates a posterior by solving an optimization problem, where a collection of particle locations is jointly updated in an iterative scheme. In this paper we use an EikoNet forward model, but this could be replaced with any other differentiable forward model. Thus, HypoSVI is a general variational approach to hypocenter inversion. We validated the method with synthetic tests and compared the locations for $\sim10k$ events in Southern California with those produced by the Southern California Seismic Network. In particular, we focused on demonstrating the reliability of the method in the presence of multi-modal posterior distributions, which SVI is well suited for handling. This is all possible because the physics informed neural network forward model is differentiable at the particle locations, which is infeasible for many conventional grid-based forward models unless they are finely meshed. We demonstrate consistent improvement over a more conventional Bayesian hypocenter method, reducing gridding artifacts and having inversion computational cost independent of the velocity model complexity.

%% file: Sections/acknowledgments.tex
\section*{Acknowledgments}
This project was partly supported by a grant from the United States Geological Survey (USGS). We would like to thank Jack Wilding and Bing Q. Li for interesting discussions about the implementation and software usage.  
We would like to thank our reviewers, Anya M. Reading and Anandaroop Ray, as well as the editor Andrew Valentine, for useful comments/corrections during the review process.

\section*{Data Availability}
The earthquake phase arrival and station locations can be downloaded from the Southern California Earthquake Data Center \hyperlink{https://scedc.caltech.edu}{https://scedc.caltech.edu}.\\
HypoSVI is available at the Github repository \hyperlink{https://github.com/Ulvetanna/HypoSVI}{https://github.com/Ulvetanna/HypoSVI}, with additional runable Colab code supplied at this Github url. The NonLinLoc control file used to generate the manuscript earthquake catalogue can be found in the Supplementary Material.

%% file: Sections/SupplementaryFigures.tex
\newpage

\appendix
\renewcommand{\thefigure}{S\arabic{figure}}

\section{Supplementary Figures}
\begin{figure*}
    \centering
    \includegraphics[width=0.8\textwidth]{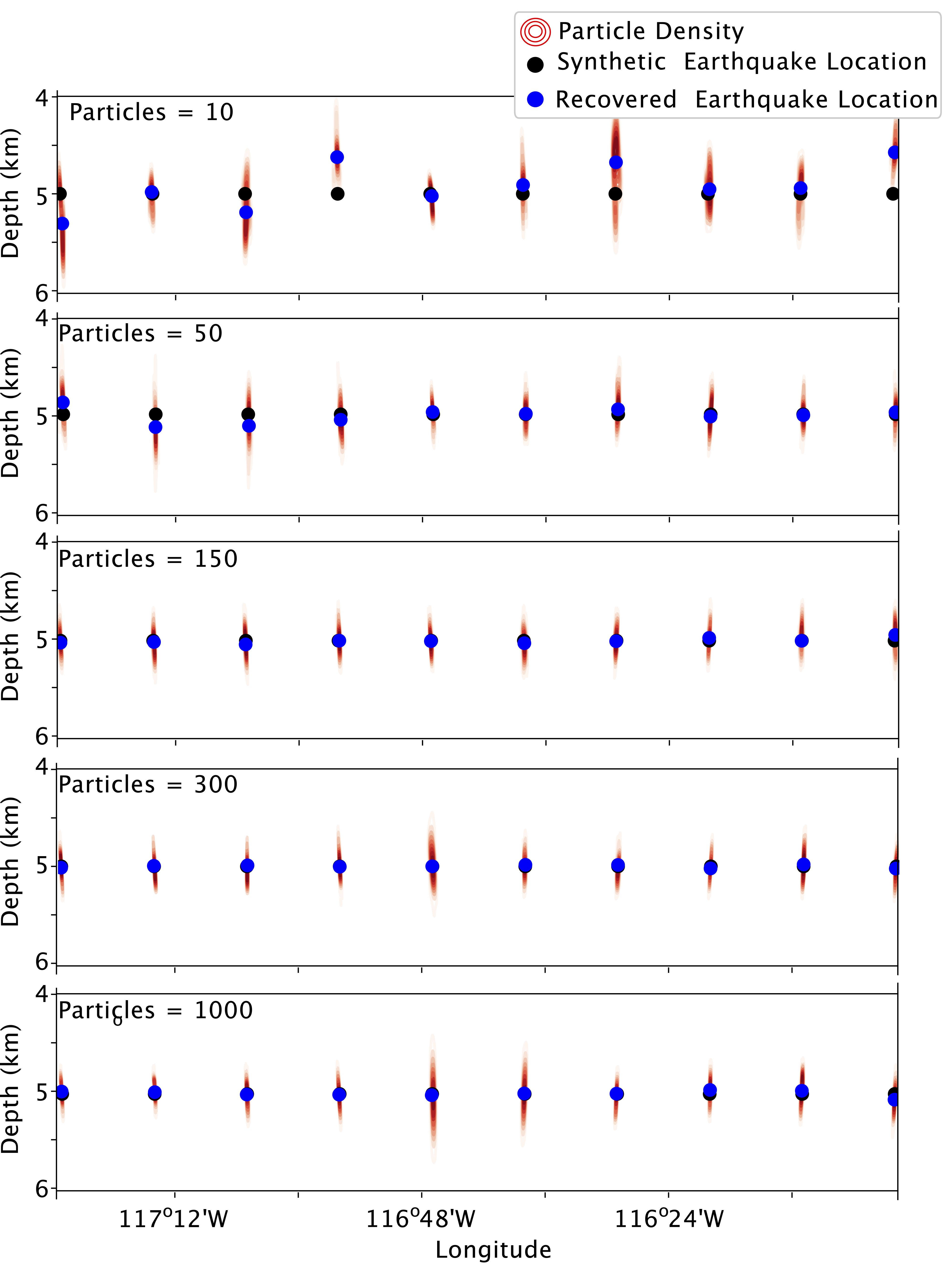}
    \caption{Synthetics earthquake location recovery for changing number of particles. An outline of observation and synthetic locations distributions is given in Section 4. Black points represent the imposed synthetic earthquake location, blue dots the recovered optimal location, red contours present the recovered posterior determined by the particle density.}
    \label{Supfig:Params_Particles}
\end{figure*}

\begin{figure*}
    \centering
    \includegraphics[width=0.8\textwidth]{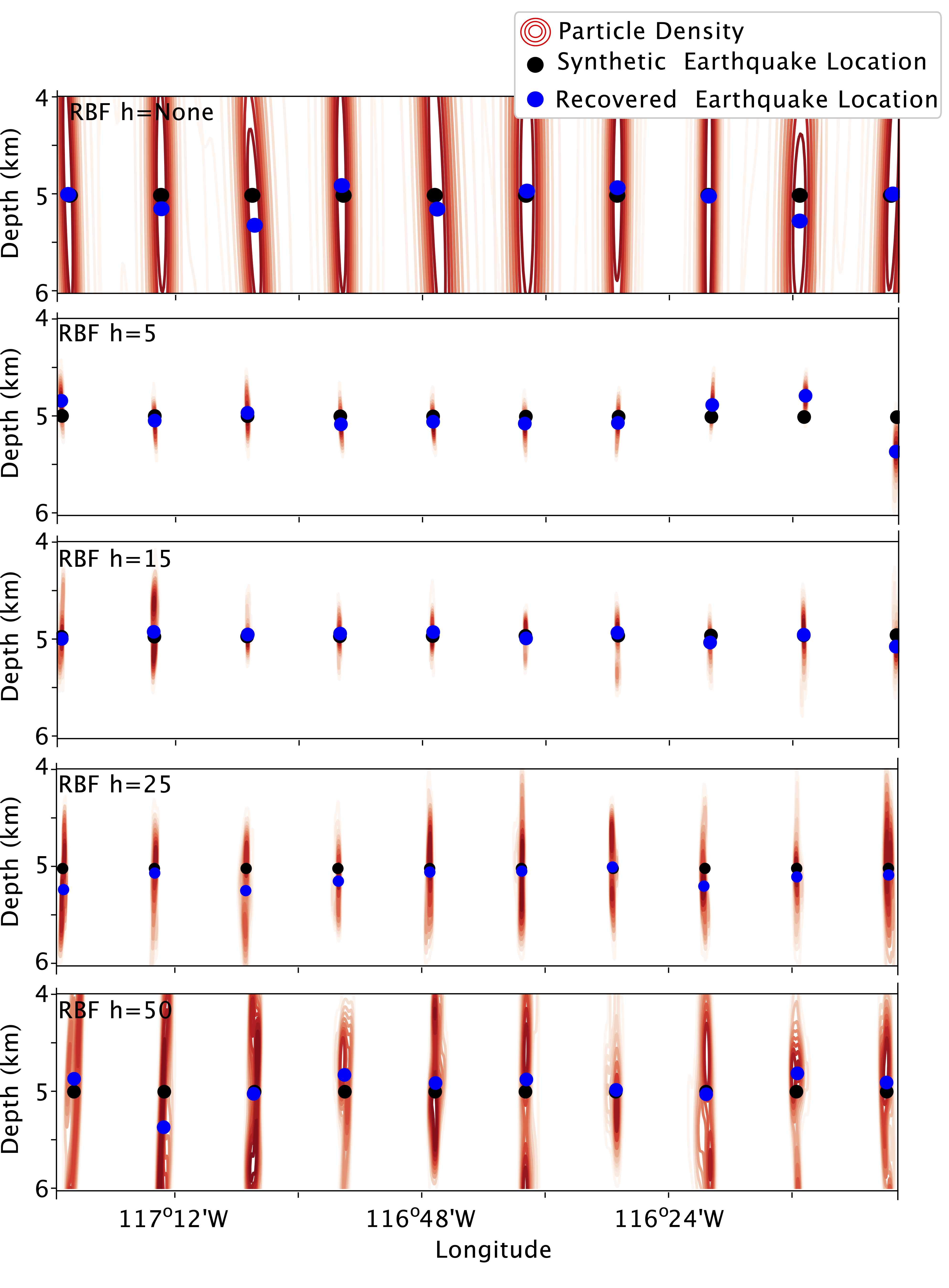}
    \caption{Synthetics earthquake location recovery for changing values for the Radial Basis Function shape parameter value.  An outline of observation and synthetic locations distributions is given in Section 4. Black points represent the imposed synthetic earthquake location, blue dots the recovered optimal location, red contours present the recovered posterior determined by the particle density.}
    \label{Supfig:Params_RBF}
\end{figure*}

\begin{figure*}
    \centering
    \includegraphics[width=0.8\textwidth]{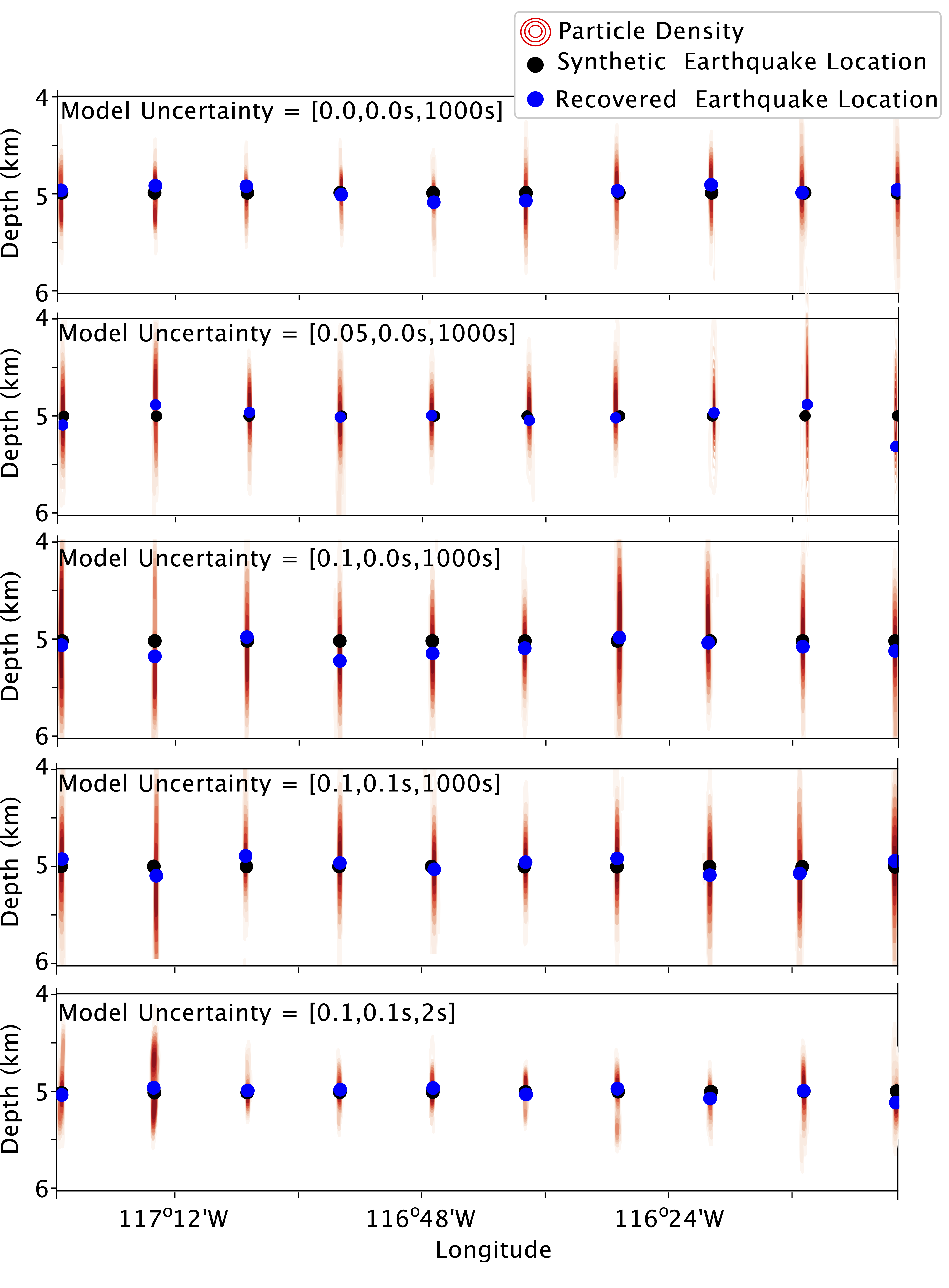}
    \caption{Synthetics earthquake location recovery for changing values for the forward model uncertainty in form $[\sigma_{f},\sigma_{min},\sigma_{max}]$.  An outline of observation and synthetic locations distributions is given in Section 4. Black points represent the imposed synthetic earthquake location, blue dots the recovered optimal location, red contours present the recovered posterior determined by the particle density.}
    \label{Supfig:Params_VeloMod}
\end{figure*}